\documentstyle[twocolumn,epsf,epsfig,aps]{revtex}

\def\lsim{\:\raisebox{-0.5ex}{$\stackrel{\textstyle<}{\sim}$}\:}
\def\gsim{\:\raisebox{-0.5ex}{$\stackrel{\textstyle>}{\sim}$}\:} 
\begin{document}
\draft
\title{
\begin{flushright}
{\small{
{TUHEP-TH-07160}\\
{SCUPHY-07001}\\
}}
\end{flushright}
\vskip -0.3cm
How to Detect `Decoupled' Heavy Supersymmetric Higgs Bosons
\vskip -0.3cm
}
\author{Mike Bisset\cite{email}, Jun Li\cite{email} 
}
\address{
Center for High Energy Physics and Department of Physics,
Tsinghua University, Beijing, 100084 P.R. China 
\vskip -0.3cm
}
\author{ Nick Kersting\cite{email}
}
\address{
Physics Department, Sichuan University Chengdu, 610065 P.R. China 
\vskip -0.3cm
}
\maketitle

\vskip -1.0cm

\begin{abstract}
Heretofore neglected decay modes of heavy MSSM Higgs bosons
into a variety of neutralino pairs may push the LHC discovery reach 
for these crucial elements of an extended Higgs sector to nearly the
TeV-scale --- if sparticle-sector MSSM input parameters are favorable.  
This is well into the so-called decoupling region \cite{decouple}, 
including moderate to low $\tan\beta$ values, where no known signals 
exist for said heavy Higgs bosons via decays involving solely 
SM daughter particles, and the lighter $h^0$ mimics the lone SM Higgs 
boson.  While the expanse of the Higgs to sparticle
discovery region is sensitive to a number of MSSM parameters, 
including in particular those for the sleptons, its presence is 
primarily linked to the gaugino inputs 
--- in fact, to just one parameter, $M_2$, if gaugino unification is 
invoked.  
Thus consideration of high {\it vs.} low $M_2$ realms in the 
MSSM should be placed on a par with the extensive consideration
already given to high {\it vs.} low $\tan\beta$ regimes.
\end{abstract}



Much effort has been devoted to finding Large Hadron Collider (LHC) 
signals for the quintet 
of Higgs bosons expected in the minimal supersymmetric standard model 
(MSSM) --- the lighter $CP$-even $h^0$, the heavier $CP$-even $H^0$, the 
$CP$-odd $A^0$, and the charged $H^{\pm}$ pair
--- and to delineating the reaches of the various signal modes
in terms of MSSM input parameters.  Key among these
have naturally been inputs affecting the tree-level
Higgs sector, $M_A$ and $\tan\beta$, and by far the 
best-known depiction of MSSM Higgs boson search projections is the 
plot of the discovery regions (d.r.'s) in the plane formed by  $M_A$ 
and $\tan\beta$.  Most studies of such signatures, including all those 
shown in well-known renditions of this plot, neglect or exclude the 
possibility of Higgs boson decays into sparticles.  The few studies
that have investigated such sparticle decay modes 
\cite{PRD1,PRD2CMS1} have focused 
almost exclusively on the four-lepton plus missing transverse energy
($E_T^{\rm{miss}}$) signature 
\begin{eqnarray} 
p p \rightarrow
H^0,A^0 \rightarrow \widetilde{\chi}_2^0 \widetilde{\chi}_2^0
\rightarrow 4\ell + 2\widetilde{\chi}_1^0 \,  
\end{eqnarray}
where $\ell=e,\mu$ and the $\widetilde{\chi}_1^0$'s,
the lightest supersymmetric particles (LSPs), 
are stable and generate the $E_T^{\rm{miss}}$.  
Previous studies have also been 
restrictive in how the $\widetilde{\chi}_2^0$'s decayed, assuming 
three-body decay modes with virtual $Z^0$'s or charged sleptons while 
neglecting  the possibility of on-shell two-body decays.

The MSSM sparticle spectrum is certainly far richer in possibilities
for Higgs boson decays, many of which can also yield the same 
$4\ell + \, E_T^{\rm{miss}}$ signature. And, as the masses of the Higgs 
bosons grow larger, ignoring such potential daughter states becomes
increasingly ill-conceived.  
The present study is a first-ever serious examination of {\em all} 
possible decay chains leading from the production of a heavy neutral MSSM
Higgs boson, $H^0$ or $A^0$, to a $4\ell +  E_T^{\rm{miss}}$ final state
\cite{Higgsslep}:
\begin{eqnarray}
p p \rightarrow
H^0,A^0 \rightarrow 
\widetilde{\chi}_a^+ \widetilde{\chi}_b^-,
\widetilde{\chi}_i^0 \widetilde{\chi}_j^0 
\rightarrow 4\ell + 2\widetilde{\chi}_1^0
\end{eqnarray}
$(a,b = 1,2, \;\; i,j = 1,2,3,4)$.
In particular, decays including the heavier neutralino (and, to a 
lesser extent, chargino) states beyond  $\widetilde{\chi}_2^0$ are found 
to dominate the signal over significant swaths of the parameter space. 

{\em Naturally}, no such signal would exist if the neutralino (and 
chargino) masses were not light enough for the Higgs bosons to decay into 
pairs of them.  Yet for the heavy (${\gsim}500\, \hbox{GeV}$) $H^0$ and
$A^0$ being considered here it is in fact {\em un-natural} for
such decay modes to be excluded.  To address this question more
quantitatively, the MSSM input parameters to the neutralino spectrum
must be incorporated.  These are, at tree-level (to which radiative
corrections are very minor\cite{inorad}): $\tan\beta$, the higgsino mixing 
parameter $\mu$, and the soft supersymmetry (SUSY)-breaking 
$SU(2)_{\hbox{\smash{\lower 0.25ex \hbox{${\scriptstyle L}$}}}}$
and
$U(1)_{\hbox{\smash{\lower 0.25ex \hbox{${\scriptstyle Y}$}}}}$
gaugino masses,
$M_{2}$ and $M_{1}$, respectively.
Gaugino unification, which draws support from the projected
merging of the  
$SU(2)_{\hbox{\smash{\lower 0.25ex \hbox{${\scriptstyle L}$}}}}$
and
$U(1)_{\hbox{\smash{\lower 0.25ex \hbox{${\scriptstyle Y}$}}}}$
coupling strengths at the GUT scale in MSSM scenarios,
imposes the TeV-scale relation
$M_{1} \simeq \frac{5}{3}\tan^2\theta_W M_{2}$ \cite{JEllis}.
Fig.\ 1 shows the inclusive production rate for 
$4\ell\,+ \, E_T^{\rm{miss}}$ events from LHC $H^0$ and $A^0$
production in the plane formed by $\mu$ and $M_2$, 
for representative values of $M_A$ and $\tan\beta$, and assuming an 
integrated luminosity of $300\, \hbox{fb}^{-1}$.  The shape of such 
contours were found to be quite insensitive to values of
$M_A \gsim 400\, \hbox{GeV}$ and $\tan\beta \gsim 5$, though rates
tended to grow with decreasing $M_A$ and increasing $\tan\beta$.
If $M_2$ is not too large, dozens to hundreds of signal events are 
expected for all values of $| \mu |$, though lower $| \mu |$ values 
are certainly more favorable.  Lower values of $| \mu |$ also enable 
large event rates for larger values of $M_2$ as shown.  
Also depicted is the percentage of signal events stemming from 
$H^0,A^0 \rightarrow \widetilde{\chi}_2^0 \widetilde{\chi}_2^0$ 
decays only.  For values of $| \mu |$ $\gsim \, 400\, \hbox{GeV}$, this 
generally exceeds $90$\%.  However, for lower values of $| \mu |$ 
and low to moderate $M_2$ values, decay modes featuring the other
heavier neutralinos,  $\widetilde{\chi}_3^0$ and/or
$\widetilde{\chi}_4^0$, become dominant.  The maxima for the signal
event rate move deeper and deeper into this previously unanalyzed
zone as $M_A$ grows large.  
Thus the optimal low  $| \mu |$ region
was overlooked by previous studies limited to only the $H^0,A^0 
\rightarrow \widetilde{\chi}_2^0 \widetilde{\chi}_2^0$
decay mode.

As the Higgs boson masses are raised and heavier neutralino
states become kinematically accessible as decay products, it
becomes ever more feasible to have slepton masses situated below the 
heavier neutralino masses.
Neutralino decays to on-shell charged sleptons
strongly enhance branching ratios (BRs) to leptonic final states,
with virtually all such decays yielding leptons (which tend to be 
stiffer due to the heavier parent neutralino).  
Thus the heavier neutralinos may well differ significantly from 
$\widetilde{\chi}_2^0$ in their capacities as conduits for the production 
of charged leptons from heavy Higgs boson production.  
Here though spoiler decay modes via on-shell sneutrinos must not
dominate, since these lead to uninteresting neutrino-bearing final states.
Fortunately this happens along relatively narrow strips in parameter 
space, such as when \cite{EWpaper}
$m_{\widetilde{\nu}} < m_{\widetilde{\chi}_2^0} < m_{\widetilde{\ell}}$,
a condition though that is quite compatible with early LHC sparticle 
detection.  
In this case the previously-studied 
$\widetilde{\chi}_2^0\widetilde{\chi}_2^0$
decay modes yield almost no signal events, while at the same parameter
space point huge event rates from the thusfar neglected heavier neutralino
channels are quite possible.

Full event generator-level studies including the simulation of a typical 
LHC detector environment (using private codes checked against results in the 
literature) were performed using HERWIG 6.5 \cite{HERWIG}
(which obtains its MSSM inputs from ISASUSY 7.58 \cite{ISAJET}) 
with the CTEQ6L \cite{CTEQ} set of parton distribution functions.
Higgs bosons were generated through the sub-processes
$q\bar q,gg\to H^0,A^0$, 
conservatively ignoring minor additional contributions from
$2 \rightarrow 3$ subprocesses which would in any event tend to fail
cuts imposed later. 
(The same HERWIG processes were used to normalize rates shown in FIG.\ 1.) 
Four-lepton events are first selected according to these criteria:
\newline
{\bf 0.}
Events have exactly four leptons, $\ell=e$ or $\mu$, 
consisting of two opposite-sign, same-flavor lepton pairs,
meeting the following criteria: 
(i) each lepton must have $|\eta^\ell|<2.4$ and
$E_T^\ell >7,4$~GeV for $e,\mu$, respectively; 
(ii) each lepton must be isolated, set by the demands that there be 
no tracks (of charged particles) with $p_T > 1.5\, \hbox{GeV}$ in a 
cone of $r = 0.3\, \hbox{radians}$ around a specific lepton, and also 
that the energy deposited in the electromagnetic calorimeter be less 
than $3\, \hbox{GeV}$
for $0.05\, \hbox{radians} < r < 0.3\, \hbox{radians}$.
\newline
Selected events are then subjected to the following cuts:
\newline
{\bf 1.}
no opposite-charge same-flavor lepton pairs may
reconstruct $M_{Z} \pm 10\, \hbox{GeV}$.
\newline
{\bf 2.}
all leptons must have
$20\, \hbox{GeV} < E_T^\ell < 80\, \hbox{GeV}$.
\newline
{\bf 3.}
events must have
$20\, \hbox{GeV} < E_T^{\rm{miss}} < 130\, \hbox{GeV}$.
\newline
{\bf 4.}
any jets (defined for $E_T > 20\, \hbox{GeV}$, $|\eta| < 2.4$ 
and with $\Delta R = 0.5$ \cite{Jets}) 
must have $E_T^{\rm{jet}} < 50\, \hbox{GeV}$.
\newline
An additional cut on the four-lepton invariant mass was also
examined; however, the optimal numerical value for this cut
depends on the masses of the Higgs bosons and the LSP, which 
are assumed to be {\it a priori} unknown.   If these masses are 
known, even approximately, then this extra cut, not employed here,
can certainly further reduce the background rates.  

All noteworthy background processes were simulated, including those 
from Standard Model (SM) processes $Z^0Z^{0(*)}$ and $t\bar tZ^{0(*)}$
and MSSM ones involving
$\widetilde{\ell}$ and/or $\widetilde{\nu}$
pair production, 
$\widetilde{\chi} \widetilde{\chi}$ or
$\widetilde{q}/\widetilde{g} \widetilde{\chi}$
pair production
(here the neutralinos and/or charginos are produced via 
an $s$-channel off-shell $Z^{0 *}$ or $W^{\pm *}$),
$\widetilde{q}$ and/or $\widetilde{g}$
pair production (here leptons are generated via 
subsequent cascade decays featuring charginos and
neutralinos, see \cite{Cascade}, but the jet cut is 
effective at eliminating such events, assuming 
reasonably heavy colored sparticles with masses
well above those of the colorless EW states), 
$t\bar th^0$, and $tH^-$ + c.c. (this last process, 
studied in \cite{charged}, could arguably be lumped
together with the signal processes; however, the cuts 
employed here are not designed to select such events,
with at best only a handful of these surviving).

Information gathered from FIG.\ 1 enables selection of  
representative MSSM input parameter sets,
aside from $M_A$ and $\tan\beta$,
for the uncolored sparticles
(colored gluinos and squarks are fixed around the TeV scale).  
For this work, one set is chosen to have
$4\ell +  E_T^{\rm{miss}}$ events coming almost exclusively from
$H^0,A^0 \rightarrow \widetilde{\chi}_2^0\widetilde{\chi}_2^0$
decays:
\newline
{\bf Set 1:}
$M_2 = 180\, \hbox{GeV}$,
$M_1 = 90\, \hbox{GeV}$,
$\mu$ = $-500\, \hbox{GeV}$,
$m_{\widetilde{\ell}_{soft}} =
m_{\widetilde{\tau}_{soft}} = 250\, \hbox{GeV}$;
\newline
while for a second set channels involving the heavier neutralinos
tend to dominate:
\newline
{\bf Set 2:}
$M_2 = 200\, \hbox{GeV}$,
$M_1 = 100\, \hbox{GeV}$,
$\mu = -200\, \hbox{GeV}$,
$m_{\widetilde{\ell}_{soft}} = 150\, \hbox{GeV}$,
$m_{\widetilde{\tau}_{soft}} = 250\, \hbox{GeV}$.
\newline
Here $m_{\widetilde{\ell}_{soft}}$ and
$m_{\widetilde{\tau}_{soft}}$ are SUSY-breaking slepton mass inputs,
with 
$m_{\widetilde{\ell}_{soft}} 
\equiv m_{\widetilde{\ell}_{\scriptscriptstyle R}}
= m_{\widetilde{\ell}_{\scriptscriptstyle L}}$,
to which the physical slepton masses are closely linked \cite{Cascade},
and $A_\tau=A_\ell=0$.
Signal(S) and background(B) rates for each parameter set are 
studied across the ($M_A$, $\tan\beta$) plane to map out 
potential d.r.'s.  The exact criteria used for demarcating d.r.'s 
are presence of at least $10$ S events and a
$99$\%-confidence-level upper limit on B events below the 
$99$\%-confidence-level lower limit on S+B events \cite{MSSMhgamgam}.
(As the number of S events grows large, this merges with the 
typical $S$/$\sqrt{B} \, \gsim \, 4$-$5$ criterion.)
Simulation runs at well over a thousand points in the ($M_A$, $\tan\beta$) 
plane yield the results shown in FIG.\ 2 and FIG.\ 3 (detailed tabular 
results from representative simulation runs will be presented elsewhere)
for {\bf Set 1} and {\bf Set 2}, respectively.
The $4\ell +  E_T^{\rm{miss}}$ d.r.'s for LHC luminosities of 
${\cal L}_{int} = 100\, \hbox{fb}^{-1}$ and $300\, \hbox{fb}^{-1}$
are delineated by the thickened contours, with the 
$100\, \hbox{fb}^{-1}$ contours inscribed within the 
$300\, \hbox{fb}^{-1}$ ones.  
Also shown are the region excluded by LEP results and the 
the expected reaches, assuming ${\cal L}_{int} = 300\, \hbox{fb}^{-1}$, 
of Higgs boson decay modes into SM daughter particles as developed by the 
ATLAS collaboration \cite{ATLASsource}. 
The new $4\ell +  E_T^{\rm{miss}}$ d.r.'s clearly provide significant 
coverage of the moderate $\tan\beta$, high $M_A$
so-called `decoupling region' \cite{decouple} inaccessible to the other 
signals \cite{ATLASsource,bbHA,htautau}.  Certainly, the 
Higgs boson to SM daghters' d.r.'s taken from the ATLAS results
are not obtained using the same choice of MSSM input parameters as are the 
$H^0, A^0 \rightarrow \widetilde{\chi}_i^0 \widetilde{\chi}_j^0$ ones 
developed in the present work.
In fact, the former d.r.'s use input choices designed to
eliminate, or at least minimize, Higgs boson decays into sparticles.
This means that, were the ATLAS reaches to be generated for
the same set of neutralino input parameters as the  $H^0, A^0
\rightarrow \widetilde{\chi}_i^0 \widetilde{\chi}_j^0$ discovery
regions, the former may well {\em shrink} in size, further emphasizing the
importance of thoroughly studying the
$H^0,A^0 \rightarrow 4\ell + E_T^{\rm{miss}}$ signature.
The diminution of the expected signatures from SM decay modes of the
MSSM Higgs bosons was investigated in \cite{PRD1,htautau}.
Assumptions inherent in the ATLAS d.r.'s for the SM decay
modes of the MSSM Higgs bosons are no less restrictive than the choices 
of MSSM input parameters made to generate the two $4\ell + E_T^{\rm{miss}}$
d.r.'s in this work.

Comparing d.r.'s for {\bf Set 1} and {\bf Set 2}, 
startling differeces are immediately apparent (these will be described in
more detail elsewhere \cite{ha4l}).  Of perhaps foremost
importance among these is {\bf Set 2}'s far greater stretch to high $M_A$ 
values.  This is due to the heavier neutralino decay modes first 
described herein:  in FIG.\ 3 corresponding to {\bf Set 2}, inclusion of 
only $\widetilde{\chi}_2^0 \widetilde{\chi}_2^0$ decay modes, as in 
previous studies, yields a mere smidgen of a d.r.\ around 
$($ $M_A$, $\tan\beta$ $)$ $=$ $($ $350$-$400\, \hbox{GeV}$, $3$-$4.5$ $)$
rather than the expansive region shown.  
All of the large portion of the d.r.\ above 
$\tan\beta = 5$ would be lost.
Note how there is some d.r.\ reach for almost all 
possible values of $\tan\beta$, 
demonstrating the relative merit of the $M_2$ and $\mu$ parameters in
demarcating heavy Higgs bosons' d.r.'s.  
Previously, particular attention to the high $\tan\beta$ regime of
the MSSM has been afforded in this journal \cite{hightb}.  It should 
however be noted that the parameter choices considered herein are no more  
limited in scope than studies devoted exclusively to that subset 
of model parameter space.  Indeed, artificially foisting sparticle
masses into the stratosphere
to avoid their presence in Higgs boson decays, as is the norm in most 
MSSM Higgs boson studies
(which then resemble two-Higgs doublet model studies)
is arguably even more restrictive.  
For all their differences, roughly representing different extremes in 
the signal composition, both {\bf Set 1} and {\bf Set 2}
offer new coverage for the crucial moderate to high $M_A$, middle
$\tan\beta$ zone.

In conclusion, restricting the search for heavy MSSM Higgs boson LHC 
signals to SM decay modes plus perhaps the single neutralino 
pair $\widetilde{\chi}_2^0 \widetilde{\chi}_2^0$ misses a significant
portion of the potential d.r.\ accessible with the inclusion of 
all sparticle decays.  A complete analysis featuring all
sparticle channels yielding the $4\ell \,+ \, E_T^{\rm{miss}}$ 
signature is quite tractable, and, for the heavy $H^0$ and $A^0$ present 
in the so-called decoupling region,
(i) this is the {\em only} known signal mode; and (ii) most of the
discovery region mapped out relies on the heavier neutralinos,
which also more easily accommodate rate-enhancing underlying
sleptonic states, never included in such analyses before.  

{\footnotesize
We are grateful to Dr. F. Moortgat for providing codes modified for
use in this project, and to Dr. S. Moretti for his critical reading of 
the manuscript.  G. Bian, Y. Liu and M. Ruan provided valuable 
assistance with graphics.
The work of MB is partially funded by a National Science Foundation of
China grant (No: 9417188).
}


\onecolumn
\centerline{}
\vskip -3.75cm
\begin{figure}[!t]
\centerline{}
\begin{center}
\epsfig{file=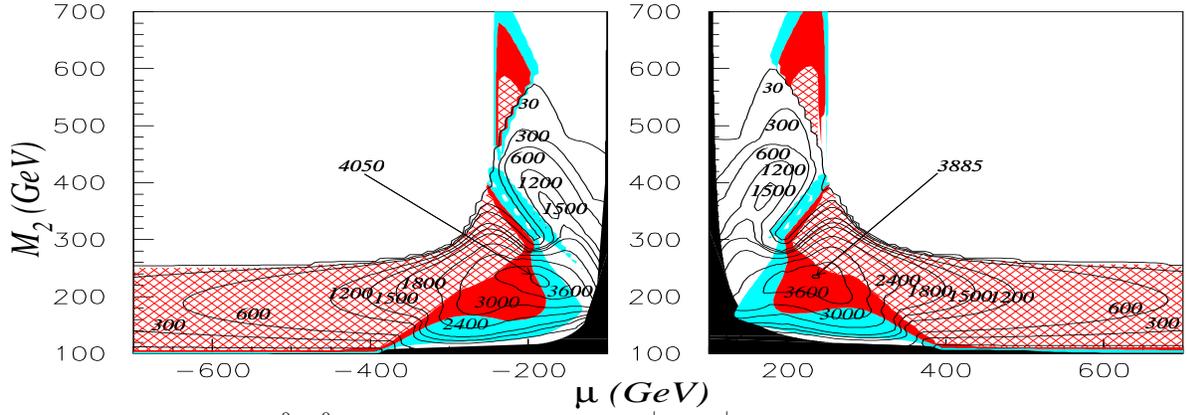,height=130mm,width=180mm}
\vskip -5.75cm
\caption{
Number of $pp \rightarrow H^0,A^0 \rightarrow 4\ell N$ events, where 
$\ell = e^{\pm}$ or
${\mu}^{\pm}$ and $N$ represents invisible final state particles, for
$300\, \hbox{fb}^{-1}$,  with
$\tan\beta = 10$ and $M_A = 500\, \hbox{GeV}$.
Also shown is the percentage from $H^0,A^0 \rightarrow
\widetilde{\chi}^0_2 \widetilde{\chi}^0_2$ 
(in color / black \& white):
$>$ 90\% (red cross-hatched / grey cross-hatched), 
50\% -- 90\% (red/dark shaded), 10\% -- 50\% 
(blue/lighter shading),
$<$ 10\% (white/no shading).
Optimized slepton masses \protect\cite{charged}
(with stau inputs raised $100\, \hbox{GeV}$)
are used, $m_t = 175\, \hbox{GeV}$, $m_b = 4.25\, \hbox{GeV}$,
$m_{\widetilde q} = 1\, \hbox{TeV}$,
$m_{\widetilde g} = 800\, \hbox{GeV}$,
$A_\tau=A_{\ell} = 0$. The blackened-out areas are excluded by LEP.}
\label{tb10ma500color}
\end{center}
\end{figure}

\vskip -1.5cm

\begin{figure}[!h]
\centerline{}
\begin{center}
\epsfig{file=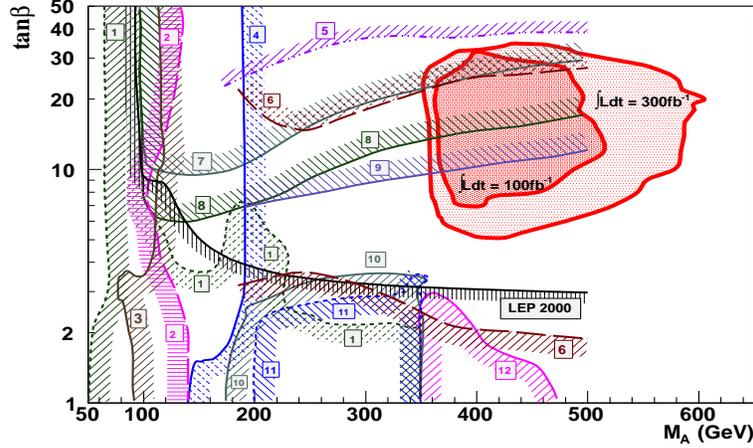,height=60mm,width=100mm}
\end{center}
\vspace{-0.2cm}
\caption{
Discovery region in ($M_A, \tan\beta $) plane
for MSSM Parameter {\bf Set 1} and 
${\cal L}_{int} = 100\, \hbox{fb}^{-1}$ \& $300\, \hbox{fb}^{-1}$
for MSSM Higgs bosons' $4\ell$ signals from their 
$\widetilde{\chi}^0_i \widetilde{\chi}^0_j,
 \widetilde{\chi}^+_m \widetilde{\chi}^-_n$ decays
(here $H^0,A^0$ decays to 
$\widetilde{\chi}_2^0 \widetilde{\chi}_2^0$ totally dominate), 
shown together with regions for other MSSM Higgs boson signatures 
from decays to SM particles based upon LEP results and 
ATLAS simulations \protect\cite{ATLASsource}
(which assume ${\cal L}_{int} = 300\, \hbox{fb}^{-1}$) where
labels represent:
1. $H^0 \rightarrow Z^0Z^{0*} \rightarrow 4\, \hbox{leptons}$;
2. $t \rightarrow b H^+, \; H^+ \rightarrow \tau^+ \nu$ + c.c.;
3. $t \bar{t}h^0,\; h^0 \rightarrow b \bar{b}$;
4. $h^0 \rightarrow \gamma \gamma$ and 
   $W^{\pm}h^0/tth^0, \; h^0 \rightarrow \gamma\gamma$;
5. $b \bar{b}H^0, b\bar{b}A^0$ with $H^0/A^0 \rightarrow b \bar{b}$;
6. $H^+ \rightarrow t \bar{b}$ + c.c.;
7. $H^0/A^0 \rightarrow {\mu}^+ {\mu}^-$;
8. $H^0/A^0 \rightarrow {\tau}^+ {\tau}^-$;
9. $g\bar{b} \rightarrow \bar{t}H+, \; H^+ \rightarrow \tau^+ \nu$ + c.c.;
10. $H^0 \rightarrow h^0h^0 \rightarrow b \bar{b}\gamma\gamma$;
11. $A^0 \rightarrow Z^0h^0 \rightarrow \ell^+ \ell^- b \bar{b}$;
12. $H^0/A^0 \rightarrow t \bar{t}$.
Note that SM discovery regions are not for 
the same input parameters: they 
presume Higgs bosons cannot decay into sparticles, so
more accurate estimates may well be smaller.
}
\label{Point1ATLAS:discovery}
\end{figure}

\vskip -1.4cm

\begin{figure}[!h]
\centerline{}
\begin{center}
\epsfig{file=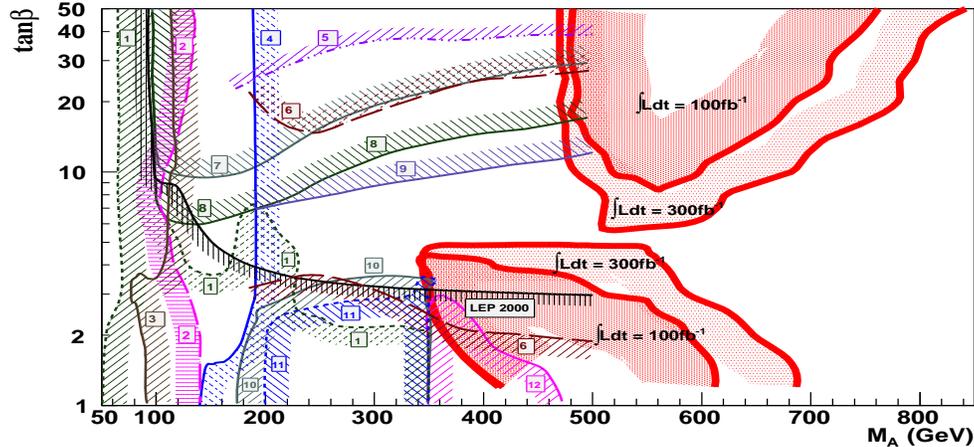,height=60mm,width=130mm}
\end{center}
\vspace{-0.2cm}
\caption{
Discovery region in ($M_A, \tan\beta $) plane for
for MSSM Parameter {\bf Set 2} and
${\cal L}_{int} = 100\, \hbox{fb}^{-1}$ \& $300\, \hbox{fb}^{-1}$
for MSSM Higgs bosons' $4\ell$ signals from their
$\widetilde{\chi}^0_i \widetilde{\chi}^0_j,
 \widetilde{\chi}^+_m \widetilde{\chi}^-_n$ decays  
(here Higgs boson decays to higher-mass neutralinos
typically dominate), shown together with regions 
for MSSM Higgs boson signatures from decays to SM particles
as in FIG.\ 2.
}
\label{Point2ATLAS:discovery}
\end{figure}


\vspace*{-0.5truecm}


\begin{thebibliography}{99}
\vspace*{-1.5truecm}


\bibitem[\ddagger]{email}
Emails: bisset@mail.tsinghua.edu.cn,   
li-j01@mails.tsinghua.edu.cn,
jx202@email.scu. edu.cn

\bibitem{decouple}
S. Moretti, Pramana {\bf 60},  369 (2003);
A.~Djouadi,
{\tt hep-ph/0503173}.

\bibitem{PRD1}
H. Baer, M. Bisset, D. Dicus, C. Kao and X. Tata,
Phys. Rev. {\bf D47}, {1062} ({1993}).

\bibitem{PRD2CMS1}
H. Baer, M. Bisset, C. Kao and X. Tata
Phys. Rev. {\bf D50}, 316 (1994);
F. Moortgat, S. Abdullin and D. Denegri,
{\tt hep-ph/0112046}; see also ATLAS TDR of \cite{ATLASsource}.

\bibitem{Higgsslep}
M. Bisset, P. Roy and S. Raychaudhuri,
{\tt hep-ph/9602430}
showed heavier MSSM Higgs boson decays to slepton pairs 
only have potentially significant BRs for
$\tan\beta \lsim \, 3$; such decays also   
generally do not yield $4\ell$ events.

\bibitem{inorad}
D. Pierce and A. Papadopoulos,
Nucl. Phys. {\bf B430}, 278 (1994),
Phys. Rev. {\bf D50}, 565 (1994).

\bibitem{JEllis}
J.R. Ellis, K. Enqvist, D.V. Nanopoulos and K. Tamvakis,
Phys. Lett. {\bf B155}, 381 (1985);

\bibitem{EWpaper}
G. Bian, M. Bisset, N. Kersting, Y. Liu, and X. Wang,
{\tt hep-ph/0611316}.

\bibitem{HERWIG}
G.~Corcella {\it et al.},   
J. High Energy Phys. {\bf 0101},  010 (2001);  
G.~Corcella {\it et al.},
{\tt hep-ph/0210213}; S.~Moretti {\it et al.},
  JHEP {\bf 0204},  028 (2002).

\bibitem{ISAJET}
H. Baer, F.E. Paige, S.D. Protopopescu and X. Tata,
{\tt hep-ph/0001086}.

\bibitem{CTEQ}
J. Pumplin {\it et al.}, 
J. High Energy Phys. {\bf 0207}, 012 (2002);
D. Stump {\it et al.},
{\it ibid.}, {\bf 0310}, 046 (2003).

\bibitem{Jets}
S.~Moretti, L.~Lonnblad and T.~Sjostrand,
J. High Energy Phys. {\bf 9808}, 001 (1998).

\bibitem{Cascade}
M.~Bisset, N.~Kersting, J.~Li, F.~Moortgat, S.~Moretti and Q.~L.~Xie,
Eur. Phys. J. {\bf C45}, 477 (2006).

\bibitem{charged}
M. Bisset, M. Guchait and S. Moretti,
Eur. Phys. J. {\bf C19}, 143 (2001);
M. Bisset, F. Moortgat and S. Moretti,
Eur. Phys. J. {\bf C30}, 419 (2003).

\bibitem{MSSMhgamgam}
H. Baer, M. Bisset, C. Kao and X. Tata,
{\bf D46},  {1067} ({1992}).

\bibitem{ATLASsource}
ATLAS TDR, Volume II, 
CERN/LHCC 99-15, May 1999;
T. Abe {\it et al.}, {\tt  hep-ex/0106056}.

\bibitem{bbHA}
J. Dai, J.F. Gunion and R. Vega
Phys. Rev. Lett. {\bf 71}, 2699 (1993),
Phys. Lett. {\bf B387}, 801 (1996);
L. Zivkovic,
CERN-THESIS-2006-063;
U. Aglietti {\it et al.}, {\tt  hep-ph/0612172}.

\bibitem{htautau}
S. Gennai {\it et al.}, {\tt  hep-ph/0704.0619}.

\bibitem{ha4l} 
M. Bisset, J. Li, N. Kersting, F. Moortgat and S. Moretti,
in preparation.
A brief synopsis:  
continuation of the {\bf Set 2} d.r.\ up beyond
$\tan\beta = 50$ while {\bf Set 1}'s dies below $\tan\beta = 35$ 
is due to elevation of {\bf Set 2} $\widetilde{\tau}$ inputs relative to 
those of $\widetilde{e},\widetilde{\mu}$.
Falling
$\Gamma(H^0, A^0 \rightarrow \widetilde{\chi}_2^0 \widetilde{\chi}_2^0)$
coupled with rising $\Gamma(H^0, A^0 \rightarrow t\bar{t})$
cut off the {\bf Set 1} d.r.\ below $\tan\beta \simeq 6$, while for 
{\bf Set 2}
$\Gamma(H^0, A^0 \rightarrow \widetilde{\chi}_2^0 \widetilde{\chi}_2^0)$
decrease less rapidly and are augmented by other
$H^0, A^0 \rightarrow \widetilde{\chi}_i^0 \widetilde{\chi}_j^0$
decays.  With {\bf Set 2}, for $\tan\beta \, \gsim \, 4.5$, 
$M_A \lsim 450\, \hbox{GeV}$, $\widetilde{\nu}$ spoiler modes kill the 
signal from 
$H^0, A^0 \rightarrow \widetilde{\chi}_2^0 \widetilde{\chi}_2^0$
and other neutralino decay modes are inaccessible.
Combined, these effects make the {\bf Set 1} and {\bf Set 2} d.r.'s
resemble mirror images of each other around
$350\, \hbox{GeV}$ to $450\, \hbox{GeV}$.

\bibitem{hightb}
H. Baer, C.-H. Chen, M. Drees, F. Paige and X. Tata
Phys. Rev. Lett. {\bf 79}, 986 (1997)


\end{thebibliography}
\end{document}